

\magnification\magstep1
\printoptions
\loadbold
\hsize=37truepc
\hoffset=4pt
\parindent=1.5em
\pageheight{45.5pc}
\pageno1

        %
meros complejos
meros enteros
meros reales

\def\hbar{\mathchar '26\mkern -9muh}   
\def\1{\'{\i}}                         

\define\bra#1{\left\langle #1 \right|} 
\define\bk#1{\left| #1 \right\rangle}  
\define\pois#1#2{\{#1,#2\}}            
\define\conm#1#2{\left[ #1,#2 \right]} 
\define\anticonm#1#2{\left\{ #1,#2 \right\} } 

\define\q#1{{\left[ #1\right]}_q}      
\define\qq#1{{\left\{ #1\right\} }_q}  

\define\op#1{\widehat{#1}}             
\define\til#1{\widetilde{#1}}          
\define\>#1{{\bold#1}}                 
n para vectores
\define\N{\Cal N}                      
\define\K{\Cal K}                      

\def\A{{\Cal A}}

\define\z#1{{\left[ #1\right]}_{q'}}      
\define\zz#1{{\left\{ #1\right\} }_{q'}}  

\def\K{{\Cal K}}

\NoBlackBoxes
\NoRunningHeads

\topmatter
\title Multiboson expansions\\
 for the $\bold q$-oscillator and ${\text{\bf su(1,1)}_q}$\endtitle
\author Angel Ballesteros and Javier Negro
\endauthor
\affil Departamento de F\1sica Te\'orica. \\
Universidad de Valladolid, 47011--Valladolid. Spain.\\
e--mail: fteorica\@cpd.uva.es
\endaffil
\email
fteorica@cpd.uva.es
\endemail

\abstract
All the hermitian representations of the ``symmetric" $q$-oscillator are
obtained by means of expansions. The same technique is applied to
characterize in a systematic way the $k$-order boson realizations of the
$q$-oscillator and $su(1,1)_q$. The special role played by the quadratic
realizations of $su(1,1)_q$ in terms of boson and $q$-boson operators is
analysed and clarified.

\endabstract \endtopmatter
\baselineskip12pt

\vfill

\pagebreak

It is well known that a way --in some sense opposite to contractions-- to
relate a pair of non--isomorphic algebras are the expansion transformations
\cite{1}.  The underlying idea of this method is that the generators of the
first algebra can be, in principle, expressed as functions of the second
ones. Only in some special cases these functions are included in the
enveloping algebra of the latter (i.e., they are polynomials) but, in
general, expansions often contain roots, rational functions or formal power
series of operators for which clear commutation rules as well as conditions
on their representation spaces must be given. A way to avoid many problems in
this respect is to work just with one representation --where expansions are
well defined-- and obtain one precise representation of the final algebra.
The general validity of the so derived relations can be studied a posteriori.

\smallskip

This method has been successfully applied in the representation theory
of quantum algebras \cite{2--3} and is known in this context as the
``deforming functional" approach \cite{4--10}.
This paper explores the possibilities of the procedure described in [9--10]
in order to build $k$--order expansions relating $su(1,1)_q$ and the
$q$--oscillator (with $q$ being always a real positive parameter). As a
byproduct, the classical relations between representations of these
algebras are obtained in the $q\to 1$ limit.

\smallskip

Section 1 begins with an application of the
expansion method that illustrates its deep connection with representations.
We ask for nontrivial expansions of the $q$-oscillator algebra
into itself. We will see that the solutions to this
problem correspond to the non equivalent irreducible
representations of this quantum algebra. These realizations are explicitly
computed, and it is shown that expansions ``intertwin" all of them.

\smallskip

In Section 2 we compute the expansions of the
$q$--oscillator $Os(1)_q$ in terms of $k$--order powers of classical boson
operators $\{N,a,a^+\}$ (the so called
$k$th order generalized $q$--boson operators \cite{11}).  The
coherent and squeezed states of this algebra have been
recently calculated \cite{11-14} and, in their non--deformed version
\cite{15}, these operators have been shown to describe non--classical
multiphoton states in quantum optics \cite{16}. The new representations of
$Os(1)_q$ found in Section 1 play a central role and allow us to classify the
solutions in their most general form. In particular, we find that the original
representation space of the classical oscillator supports $k$ independent
irreducible representations of $Os(1)_q$ which are explicitly given.

\smallskip

Section 3 carries out the
same program for realizations of $su(1,1)_q$ in terms of $k$--order
boson operators. There are several remarkable results:
Firstly, we get also $k$ independent representations in a similar way to the
oscillator case. Secondly, it is possible to obtain in this way any discrete
series representation of $su(1,1)_q$  (both positive $l>0$ and
negative $l<0$). Finally, it is shown that only for the non--deformed $k = 2$
case it is possible to define expansions within the enveloping algebra. In
particular, for $l = -1/4$ we find the realization given by
$$
K_3 = -\tfrac12 (N +\tfrac12), \quad K_+ =
\tfrac12 a^2,\quad K_- = \tfrac12(a^+)^2.
$$
This is a discrete negative series
dual of a well known positive relation that holds both in quantum and
classical cases \cite{17--18}. The ``quantum" part of this statement is
proved in Section 4, where all $q$-boson realizations of $su(1,1)_q$  are
derived. As a result, second order expansions are shown again to be the only
ones that can be  written in the abovementioned form. Some comments in the
final section will end the paper. \bigskip \medskip

\heading {1. The Expansion Method and $q$--Oscillator Representations}
\endheading
\medskip

We shall study the representations of the $q$--oscillator \cite{19-20} algebra,
given by the commutators:
$$
[\N,\A^+] = \A^+,\quad [\N,\A] = -\A,\quad [\A,\A^+] = \{\N +
\tfrac12\}_q,\tag1.1
$$
where $\qq{x} = \dfrac {q^x +q^{-x}}{q^{\frac 1 2} + q^{-\frac 1 2}}$. This
definition means that we have a ``symmetric" $q$--oscillator: $q$ and $q^{-1}$
give the same commutation relations. Representations for other non
symmetric cases have been studied in \cite{21--22}.

\medskip

Firstly we characterize
the representations where $\N$ is a hermitian operator, while $\A^+$ is the
hermitian conjugated of $\A$. It is possible to prove by a lengthy but
straightforward computation that $C_q = [\N]_q -\A^+\A$ is a Casimir
operator for this algebra (here the standard notation
$\q{x} = \dfrac {q^x -q^{-x}}{q - q^{-1}}$ has been used). Let
$|\lambda\rangle$ be an eigenstate of $\N$ with eigenvalue $\lambda$. Then,
making use of the commutations (1.1) we can see that $(\A^+)^k|\lambda\rangle$,
if non-null,  will be an $\N$ eigenvector with eigenvalue $\lambda + k$, while
$(\A)^{k'}|\lambda\rangle$ will have eigenvalue $\lambda - k'$. Let
$|\mu\rangle$ be a non-null eigevector, $\N|\mu\rangle = \mu|\mu\rangle$,
obtained from $|\lambda\rangle$ in this way. We compute
$$
\eqalign{
\langle\mu|C_q|\mu\rangle &= c_q \cr &=\langle\mu|[\N]_q -
\A^+\A|\mu\rangle= [\mu]_q - \langle\mu|\A^+\A|\mu\rangle.}
\tag1.2
$$
Since $\q{\mu} = \dfrac {q^\mu -q^{-\mu}}{q - q^{-1}}$, where $\mu$ and $q$
($>0$) are real numbers, then $[\mu]_q \to -\infty$ if $\mu \to -\infty$.
Taking into account that $\langle\mu|\A^+\A|\mu\rangle$  is positive and, in
order to be consistent with the first equation of (1.2), it is necessary that
some minimun value $k_1$ such that $(\A)^{k_1}|\lambda\rangle = 0$ exists.
Let us call $|0\rangle$ the normalized $\N$ eigenstate with lowest $\nu$

eigenvalue, i.e., $\N|0\rangle = \nu \bk{0} $. The whole representation space
can be generated out of this vector by applying $\A^+$ repeatedly. Hereafter,
we shall call this the $\nu_q$--representation. It is easy to show that the
underlying Hilbert space is just the number state space  ${\Cal H}$. The
explicit action on the $\{|n\rangle\}$--basis can be computed in a standard
way and reads  $$ \eqalign{ \A^+_\nu|n\rangle &= \sqrt{[\nu+n+1]_q -
[\nu]_q}|n+1\rangle \,\cr  \A^+_\nu|n\rangle &= \sqrt{[\nu+n]_q -
[\nu]_q}|n-1\rangle \,\cr \N_\nu|n\rangle &= (n+\nu)|n\rangle.}\tag1.3
$$
The Casimir eigenvalue for this representation is $c_q = [\nu]_q$. When $\nu =
0$ we have the standard ``$n_q$--representation" of $Os(1)_q$ \cite{19-20}.

\medskip

Now, we shall see how expansions can be used to get these
representations, as well as the $q$--Casimir starting from the
usual $n_q$ realization. Let us take
$$
\aligned
\A_\nu^+ &= \A^+f^+(\N),\quad \\
\A_\nu &= f^-(\N)\A,\quad \\
\N_\nu &= \N+\nu,
\endaligned
\tag1.4
$$
where $f^+(\N)$,$f^-(\N)$ are functions to be determined and $\A^+$, $\A$, $\N$
are the operators of the standard representation $n_q$, that act in the
following way:
$$
\A|n\rangle = \sqrt{[n]_q}|n-1\rangle,\quad
\A^+|n\rangle = \sqrt{[n+1]_q}|n+1\rangle,\quad
\N|n\rangle = n|n\rangle. \qquad n=0,1,2\dots\tag1.5
$$
\medskip

We want that the operators defined by the Ansatz (1.4) reproduce the same
algebra given by (1.1). All that means that in fact we are looking for
expansion transformations inside the same
algebra. By imposing (1.1) on
the basis vectors the following recurrence relation is originated:
$$
[n+1]_qf^+(n)f^-(n) - [n]_qf^+(n-1)f^-(n-1) = \{n + \nu +\tfrac12\}_q.\tag1.6

$$
The general solution of (1.6) depends on the arbitrary parameter $\nu$ and has
the
following form:
 $$
\Lambda(n_q|\nu_q) \equiv f^+(\N)f^-(\N)  = \frac{[\N+\nu+1]_q -
[\nu]_q}{[\N+1]_q}.\tag1.7
$$
The operator $\Lambda(n_q|\nu_q)$ that displays an infinite number of
expansion solutions (note that (1.7) defines only the product $f^+(\N)f^-(\N)$)
will be called the ``characteristic functional" \cite{9}. The arguments of
$\Lambda$ are the original and final representation, respectively.  The
hermiticity relation between $\A$ and $\A^+$ selects one particular
realization:
$$
f^+(\N) = f^-(\N) = \sqrt{ \frac{[\N+\nu+1]_q -
[\nu]_q}{[\N+1]_q}}.\tag1.7$^\prime$
$$
Now if we let the operators (1.4) with the functions $f^\pm$ defined by
(1.7$'$)
act on the basis $\{|n\rangle\}$ by means of (1.5), we get all the
representations (1.3) previously found. In order to deal with the Casimir
operator, we start with the usual representation characterized by $C_q = [\N]_q
-\A^+\A = 0$. Then, if we substitute expressions (1.4) we will find
$$
[\N_\nu - \nu]_q - \Lambda(n_q|\nu_q)^{-1}\A_\nu^+\A_\nu = 0, \tag1.8
$$
that taking into
account (1.7)  gives $[\N + \nu]_q - [\nu] - \A_\nu^+\A_\nu = 0$, or better:
$$
[\N _\nu]_q - \A_\nu^+\A_\nu = [\nu]_q,\tag1.8$'$
$$
as it should be.

\medskip

Observe that in the limit $q \to 1$ the  $\nu_q$--representations (1.3)
of the $q$--oscillator turn into the corresponding ones for the classical
oscillator:
$$
\eqalign{
a^+|n\rangle &= \sqrt{n+1}\,|n+1\rangle , \cr
a|n\rangle &= \sqrt n \,|n-1\rangle ,\cr
N|n\rangle &= (n+\nu)\,|n\rangle.}\tag1.9
$$
We shall see in the next sections that the
representations here analized will appear quite naturally when we study the
links between the quantum algebras $su(1,1)_q$ and $Os(1)_q$ by using
arbitrary powers of their raising and lowering operators.

\bigskip
\medskip
\heading {2. $k$--Order $q$--Oscillator Realizations}
\endheading
\medskip

We start from the usual representation $\nu =
0$ of $Os(1)$ in order to get the commutation rules of $Os(1)_q$. In this case
we formulate the expansion in the following way:

$$
\eqalign{
\N &=\tfrac1 k (N + p ),\cr
{\Cal A}^+ &=(a^+)^k g_k^+(N),\cr
{\Cal A} &=g_k^-(N) \ a^k.}
\tag 2.1
$$
We immediately see that, with this definition, the original representation
space of $Os(1)$ gives rise to $k$ different invariant subspaces. The
highest weight vector
for each expansion is, respectively, $|0\rangle$, $|1\rangle$, \dots,
$|k-1\rangle$. The whole support space for these $k$ representations is
generated by applying the operators (2.1) on the corresponding fundamental
state.

We impose the commutation rules (1.1) to hold on every basis vector
$\bk{n}$. Taking into account (2.1) we obtain
$$
{{n!}\over{(n-k)!}}g_k^+(n-k)g_k^-(n-k) - {{(n+k)!}\over
{n!}}g_k^-(n)g_k^+(n)=\qq{{1\over k}\left(n + p\right) + \tfrac 1 2}.
\tag 2.2
$$
This recurrence can be solved (see [9]) within each $m$--subspace generated by
$\bk{n}$ if $n = ks + m$, with $s$ any positive integer number and $m$
fixed such that $0 \le m < k$ .  Each $m$--subspace support an
expansion characterized by an arbitrary value $p$ that will be denoted by
$p_m$,
$m=0, 1, 2,\dots,k-1$. Thus, we obtain the functional $\Lambda_k^{p_m} (n |
m_q)=g_k^+(n)g_k^-(n)$  whose matrix elements read:
$$
\aligned
\Lambda_k^{p_m} (n | m_q)&={{n!}\over{(n+k)!}}\bigg\{ \qq{\tfrac 1 k (n + p_m)
+
\tfrac 1 2}  +\qq{\tfrac 1 k (n -k + p_m) + \tfrac 1 2} \\
&\qquad +\qq{\tfrac 1 k (n -2k + p_m) + \tfrac 1 2} +\dots
+\qq{{\tfrac 1 k}(m+p_m) + \tfrac 1 2}\bigg\}\\
&= {{n!}\over{(n+k)!}} \q{\tfrac 1 {2k}  (n-m) +
\tfrac 12} \qq{\tfrac 1 {2k} (n +m+2p_m) + \tfrac 12}(q^{1/2} + q^{-1/2}).
\endaligned
\tag 2.3
$$

We shall write (2.3) in
another form that will be more appealing for its interpretation:
$$
\Lambda_k^{p_m} (n | m_q)=
\frac{1}{(N+1)\cdots(N+k)}\left(\left[\tfrac1k(N+p_m)+1\right]_q
-\left[\tfrac1k(m+p_m)\right]_q\right) .\tag2.4
$$
If we rename $N_m = \frac1k(N-m)$ and $\nu_m = \frac1k(m+p_m)$, this
expression turns into
$$
\Lambda_k^{p_m} (n | \nu_{m(q)})=
\frac{1}{(N+1)\cdots(N+k)}\left(\left[N_m+\nu_m+1\right]_q
-\left[\nu_m\right]_q\right) .\tag2.5
$$
With this notation, a hermitian solution for each  $m$ representation
space is
$$
\eqalign{
\N_m &= N_m + \nu_m\cr
\A^+_m &=
(a^+)^k\sqrt{\frac{1}{(N+1)\cdots(N+k)}\left(\left[N_m+\nu_m+1\right]_q
-\left[\nu_m\right]_q\right)},\cr \A_m &=

\sqrt{\frac{1}{(N+1)\cdots(N+k)}\left(\left[N_m+\nu_m+1\right]_q
-\left[\nu_m\right]_q\right)}\, a^k }.\tag2.6 $$
It is clear from (2.6) and (1.3) that the $m$--subspace supports a
representation similar to one of those deduced in Section 1 and
characterized by the value $\nu_m = \frac1k(m+p_m)$. This method gives
rise to $k$ independent (infinite) families of realizations of $Os(1)_q$ (note
that we obtain any $\nu_q$ representation of $Os(1)_q$, and not only the
standard ``$n_q$'').

\medskip

Remark that, since $p_m$ is an arbitrary real number, we can
choose $p_m = -m$, for $m = 0,1,\dots k-1$. In this way we get $\nu_m = 0,
\forall m$. We then have the $k$ order generalization of the standard
representation of the $q$--oscillator $\nu_q = \q{0}$, whose
hermitian expansion has the form
$$
\eqalign{
\N_m &= N_m\cr
\A^+_m &= (a^+)^k \sqrt{\frac{1}{(N+1)\cdots(N+k)}\left[N_m+1\right]_q}\cr
\A_m &=  \sqrt{\frac{1}{(N+1)\cdots(N+k)}\left[N_m+1\right]_q}(a)^k. }
\tag2.7
$$
If we compare (2.7) with expression (11) given in \cite{11}
(where it is used the notation $\{b,b^{+}\}$, instead of our present
$\{a,a^{+}\}$), we realize
that both are formally identical. However, the generalized number operator
$N_m$ has now an explicit form in terms of $N$ that depends on the
selected subspace and avoids the use of the integer part function given in
\cite{11}. In particular
for $k = 2$ we have two hermitian realizations of the standard ``$n_q$''
representation: $$
(m=0) \quad \left\{ {\eqalign{\N &= \tfrac12N,\cr
\A^+ &= (a^+)^2\sqrt{\frac1{(N+1)(N+2)}[\tfrac12N+1]_q},\cr
\A&=\sqrt{\frac1{(N+1)(N+2)}[\tfrac12N+1]_q}\,a^2. }}\right.\tag2.8a
$$
$$
(m=1) \quad \left\{ {\eqalign {\N &=\tfrac12N-\tfrac12,\cr

\A^+&=(a^+)^2\sqrt{\frac1{(N+1)(N+2)}[\tfrac12N+\tfrac12]_q},\cr
\A &=\sqrt{\frac1{(N+1)(N+2)}[\tfrac12N+\tfrac12]_q}\,a^2. }}\right.
\tag2.8b
$$

When $q\to 1$ these
two realizations turn into
$$
\eqalign{
&i)\quad \N = \tfrac12N,\quad \A^+ = (a^+)^2\sqrt{\frac1{2(N+2)}},
\quad \A = \sqrt{\frac1{2(N+2)}}\,a^2 \cr
&ii)\quad \N = \tfrac12N-\tfrac12,\quad \A^+ = (a^+)^2\sqrt{\frac1{2(N+1)}},
\quad \A = \sqrt{\frac1{2(N+1)}}\, a^2. }\tag2.9
$$

\bigskip
\medskip

\heading 3. $su(1,1)_q$ Boson Realizations of Order $k$
\endheading
\medskip

The quantum deformation of $su(1,1)$ has the following commutation rules
\cite{17}:
$$
\eqalign{
[\K_3,\K_\pm] &= \pm\K_\pm,\cr
[\K_+,\K_-] &= -[2\K_3]_q .\cr}
\tag 3.1
$$
Its discrete series representations are labelled by $l$, the
eigenvalue of $\K_3$ on the highest weight vector. The Casimir operator is
given by
$$
C_q = [\K_3 - {\tfrac 1 2}]_q^2 - \K_+\K_-=\q{l-{\tfrac 1 2}}^2
\tag 3.2
$$
Discrete series representations of $su(1,1)_q$ have been studied from an
expansion point of view in \cite{9--10}. Now we take the classical oscillator
and its $\nu = 0$ representation space, and we define the deformed generators
of $su(1,1)_q$ as follows (see also, in this respect, an equivalent formulation
in \cite{11}) $$
\eqalign{
\K_3 &=\tfrac1 k (N + p ),\cr
\K_+ &=(a^+)^k g_k^+(N),\cr
\K_- &=g_k^-(N) \ a^k.\cr}
\tag 3.3
$$
(We have to include the $\tfrac 1 k$ factor to preserve the
commutation relations
$\conm{\K_3^k}{\K_\pm^k}=\pm\K_\pm^k$ $(k=1,2,\dots)$).
By imposing the last commutation rule in (3.1) on the operators defined by
(3.3), we obtain
$$
{{n!}\over{(n-k)!}}g_k^+(n-k)g_k^-(n-k) - {{(n+k)!}\over
{n!}}g_k^-(n)g_k^+(n)=- \q{{2\over k}\left(n + p\right)}.
\tag 3.4
$$

We get again a recurrence relation where $n = ks + m$, with $k,s,m$
integers and $0 \le m < k$.
Hence,  there will be $k$ independent expansion constants, $p_m$. A
straightforward computation shows that the matrix elements of the

characteristic functional $\Lambda$, in the oscillator basis $\{|n\rangle\}$,
are
 $$
\Lambda_k^{p_m}(n | l_q)=g_k^+(n)g_k^-(n) = {{n!}\over{(n+k)!}}
\q{\tfrac 1 k  (n +2p_m + m)}
\q{\tfrac 1 k (n - m +k)}.
\tag 3.5
$$

In conclusion, we can say that the original representation of the
classical harmonic oscillator gives rise to a reducible representation of the
quantum algebra $su(1,1)_q$ that can be decomposed into
$k$ irreducible representations --each of them labelled by $p_m$--
by means of creation and anihilation operators of degree $k$. We shall study
some cases separately.

\medskip
\subheading{3.1. First order realizations}
\smallskip

If $k=1$, then $m=0$. Any representation $l \equiv p_0(>0)$ is
realizable and we obtain:
$$
\Lambda_1^l(n | l_q)=g_1^+(N)g_1^-(N) = {{1}\over{(N+1)}} \q{ N
+2l}\q{N+1}.
\tag 3.6
$$
The classical limit of (3.6) is $\lim_{q \to 1}\Lambda_1^l(n | l_q) = (N +2l)$
\cite{9} and relates the usual representation of the harmonic oscillator with
any $l$--representation of the Lie algebra of $su(1,1)$
(for instance, Holstein--Primakoff realizations \cite{23} are included within
this class).

\bigskip
\subheading{3.2. Quadratic realizations}
\smallskip

Let us take $k=2$. We have two options ($m=0,1$),
that originate the following characteristic functionals for
$p_0, p_1$:
$$
\align
\Lambda_2^{p_0}(n | l_q) &= {{1}\over{(N+2)(N+1)}} \q{\tfrac 1 2 (N
+2p_0)}
\q{\tfrac 1 2 (N +2)}, \quad (m=0),\tag 3.7a\\
\Lambda_2^{p_1}(n | l_q) &= {{1}\over{(N+2)(N+1)}} \q{\tfrac 1 2 (N
+2p_1  + 1)}
\q{\tfrac 1 2 (N+1)}, \quad (m=1).\tag 3.7b
\endalign
$$
Both functionals coincide only when $p_0=\tfrac 1 2$ and $p_1=\tfrac 1 2$.
In that case, the corresponding irreducible representations of $su(1,1)_q$
are $l_0 = \frac 14$ and $l_1 = \frac 34$, and the unique functional
 is given by
$$
\Lambda_2 (n | l_q)=g_2^+(N)g_2^-(N) = {{1}\over{(N+2)(N+1)}}
\q{\tfrac 1 2 (N +1)}
\q{\tfrac 1 2 (N +2)}.
\tag 3.8
$$

An hermitian realization ($g_2^+(N)=g_2^-(N)$) will act on the
number state space in the form
$$
\eqalign{
\K_3\bk{n}&=\tfrac 1 2 (n +
\tfrac 1 2)\bk{n},\cr
\K_+\bk{n}&=\sqrt{{\q{\tfrac 1 2 (n+1)}}{\q{\tfrac 1
2 (n+2)}}}\bk{n+2},\cr
\K_-\bk{n}&=\sqrt{{\q{\tfrac 1 2 n}}{\q{\tfrac 1 2
(n-1)}}}\bk{n-2}.\cr}
\tag 3.9
$$

The classical limit $q \to 1$ of (3.7 a--b) is
$$
\align
\Lambda_2^0(n | l)  &= {{N + 2p_0}\over{4(N+1)}} , \quad
(m=0),\tag 3.7$'$a\\
\Lambda_2^1(n | l)  &= {{N + 2p_1 + 1}\over{4(N+2)}}, \quad
(m=1).\tag 3.7$'$b \endalign
$$
Moreover, it is clear that if $p_0 = p_1 = \frac 12$ both solutions
(3.7$'$a--b) give
the well known classical quadratic realizations of
$su(1,1)$ associated to the representations $l_0 = \frac14, l_1 = \frac 34$
\cite{11,17}:
$$
K_3=\tfrac1 2 (N + \tfrac1 2 ),\qquad
K_+=\tfrac1 2 (a^+)^2 ,\qquad
K_-=\tfrac1 2 a^2.
\tag 3.10
$$

Then, what is so special about the $l=\tfrac 1 4,\ \tfrac 3 4$
representations? They are the only ones where the quotients in (3.7$'$a--b)
equals 1 and $g_2^+(N) (= g_2^-(N) = 1)$ can be chosen within the oscillator
enveloping algebra. Any other $l$--representation of
$su(1,1)$ can be obtained from the ``$n$" representation of $Os(1)$
provided that more general functions $g_2^\pm$ are allowed.

\bigskip
\subheading{3.3. The general case $k > 2$}
\smallskip

We take the following notation: $l_m = \dfrac {p_m + m}k$ and $N_m =
\dfrac {N-m}k$. Then, within each irreducible subspace
${\Cal H}_m$, the  operator $N_m$ takes integer values 0, 1, 2\dots and we
have
$$
\Lambda_k^{l_m}(n | l_q) = {{1}\over{(N+1)\cdots(N+k)}} \q{
N_m +2l_m}
\q{N_m + 1}.
\tag 3.11
$$
Note that this expression resembles strongly the $k=1$ case.

\medskip

The Casimir operator of $su(1,1)$ for each subspace ${\Cal H}_m$ reads
$$
\q{\K_3 - \tfrac 12}^2 - \K_+\K_- = \q{l_m - \tfrac 12}^2.\tag3.12
$$

\bigskip
\subheading{3.4. Negative discrete series representations}
\smallskip

Upper bounded ( $l<0$ ) discrete representations can
be constructed in the same way. In order to see this explicitly let us define

$$
\eqalign{
\K_+ &= a^kg_k^+(N),\cr
 \K_- &= g_k^-(N) (a^+)^k,\cr
\K_3 &= -\tfrac 1k(N  + p_m).}\tag3.13
$$
In this case the matrix elements of the characteristic functional are given
by
$$
\Lambda_k^{p_m}(n|l_q) = \frac{(n-k)!}{n!}
\left[\tfrac1k(n+m+2p_m)-1\right]_q\left[\tfrac1k(n-m)\right]_q,\tag3.14
$$
where $m$ and $p_m$ have the same meaning as in the previous case.
If $k=2$, only for the values $p_0 = \frac12$, $p_1 = \frac 12$, (the
representations of $su(1,1)_q$ are $l_0=-\frac14$ and $l_1 = -\frac 34$
respectively), both functionals coincide and take the form
$$
\Lambda_2^{p_m}(n|l_q) = \frac 1{N(N-1)}
\left[\tfrac12N\right]_q\left[\tfrac12(N-1)\right]_q.\tag3.15
$$
In the classical limit $q \to 1$, (3.15) gives the usual
$N$--independent expression: $\Lambda_2(n|l_q)  = \frac 14$. However, the
realization of the $su(1,1)$ algebra is slightly different than (3.10); here
we have
$$
K_3=-\tfrac1 2 (N + \tfrac1 2 ),\qquad
K_+=\tfrac1 2 a^2 ,\qquad
K_-=\tfrac1 2 (a^+)^2.\tag3.16
$$

\bigskip
\medskip

\heading 4. $q$--Boson Realizations
\endheading
\medskip

In the previous section, we discussed the realizations of $su(1,1)_q$ as
$k$--order powers of classical bosons. Here we will address the same
question but using $q'$-bosons. It must be remarked that, in principle,
$q$ and $q'$ are independent deforming parameters.

\medskip

We can solve this problem by means of the characteristic functional
method taking into account results already obtained in the previous
sections. As a first step, we recall the expression (3.3) where the
$su(1,1)_q$ generators are written as $k$-order bosons
$$
\eqalign{
\K_3 &=\tfrac1 k (N + p ),\cr
\K_+ &=(a^+)^k g_k^+(N),\cr
\K_- &=g_k^-(N) \ a^k.\cr}
\tag 4.1
$$
The associated functional $\Lambda_k^{p_m} (n | l_q)$ is given by (3.5).
Afterwards, we can substitute these classical bosons in terms of the $q'$
bosons by inverting (2.1) with $k=1$ and $p=0$:
$$
\aligned
N&=\N,\\
a^+ &= \A^+ f_+(N), \\
a &= f_-(N)\A.
\endaligned
\tag 4.2
$$
The resultant
expansion is given by the characteristic functional $\Lambda(n_s|
n)=f_+(N)\ \! f_-(N)={{(N + 1)}\over{\z{N+1}}}$ \cite{9}.
It is easy to check that we
can generalize this statement to the $k$--order case by writing
$$
\aligned
(a^+)^k &= (\A^+)^k f_+(N)\ \! f_+(N + 1)\dots f_+(N +k - 1), \\
(a)^k &= f_-(N)\ \! f_-(N + 1)\dots f_-(N +k - 1) \  (\A)^k.
\endaligned
\tag 4.3
$$
This expansion is characterized by
$$
\Lambda_k(n_s|n)={{(N + 1)}\over{\z{N+1}}}
{{(N + 2)}\over{\z{N+2}}}\dots{{(N + k)}\over{\z{N+k}}}.
\tag 4.4
$$

Finally, we substitute (4.3) in (4.1) to get the complete expansion. The
global functional is the product of the expansion components (3.5) and (4.4):
$$
\aligned
\Lambda_k^{p_m} (n_{q'} | l_q)&= \Lambda_k (n_{q'} | n)\, \Lambda_k^{p_m}
(n | l_q)\\
&={{(N + 1)}\over{\z{N+1}}}
{{(N + 2)}\over{\z{N+2}}}\dots{{(N + k)}\over{\z{N+k}}} {{\q{\tfrac 1 k  (N
+2p_m + m)} \q{\tfrac 1 k (N - m +k)}}\over{(N +1)(N + 2)\dots (N + k)}}.
\endaligned
\tag 4.5
$$

Since both (3.5) and (4.4) contain two $q$-numbers,  it is easy to see that
the case $k=2$ is the only one whose functional could be, in principle,
simplified. We have already shown that --just for this case--
$\Lambda_k^{p_m}(n | l_q)$ can be taken independent of $m$ (3.8). With this
prescription,
$$
\Lambda_2 (n_{q'} | l_q)= {{\q{\tfrac 1 2  (N +1)}
\q{\tfrac 1 2 (N +2)}}\over{\z{N+1}\z{N+2}}}.
\tag 4.6
$$

If $q^2=q'$, (4.6) turns into $\Lambda_2 (n | l_q)=(q + q^{-1})^{-2}$, that
corresponds to the known quadratic $q$--boson realization given by Kulish
and Damashinski \cite{17}:
$$
\K_3=\tfrac1 2 (\N + \tfrac1 2 ),\qquad
\K_+=\tfrac1 2 (\A^+)^2 ,\qquad
\K_-=\tfrac1 2 (\A)^2.
\tag 4.7
$$
We remark that with (4.7) we obtain two different $su(1,1)_q$
representations ($l=\tfrac 1 4$ and $l=\tfrac 3 4$) depending on the
number subspace we are acting on (this fact was already indicated in
\cite{17}). Moreover, this representation dependence is enhanced by
the fact that this relation is not valid if we start with $\nu_m\neq 0$
representations of the $q'$--bosons. On the other hand, the same
characterization (4.6) can be reproduced for the quadratic negative series
expansion (3.15), obtaining another Kulish--Damashinski type realization:
$$
\K_3=-\tfrac1 2 (\N + \tfrac1 2 ),\qquad
\K_+=\tfrac1 2 \A^2 ,\qquad
\K_-=\tfrac1 2 (\A^+)^2.\tag4.8
$$

\medskip

The same composition of expansions can
be used to obtain all $q$--generalized boson realizations.
If we take into account (2.7), this expansion leads us to
replace the functional (4.4) by the inverse of (2.5) (with
$\nu_m=0$). A straightforward computation shows that the analogue of
(4.5) is
$$
\Lambda_k^{p_m} (n_q^k | l_q)=\q{\N_m +2l_m}.
\tag4.9
$$
To compare this infinite family of realizations with \cite{11} we have to
recall the considerations done in Section 3. Functional (4.9) contains
Holstein--Primakoff and Dyson realizations used in \cite{13}.

\medskip

To end with, we emphasize that the technique we have outlined here is
quite general and can be applied to many other algebras. For instance,
$k$-order realizations of $su(2)_q$ (either in terms of ($q$-)boson
operators or starting from $su(1,1)$ discrete series) can be
obtained (as has been done in Ref\. \cite{11}). Some other interesting problems
which are in progress arise if one wants to generalize the characteristic
functional approach to simple quantum algebras with higher rank.

\bigskip
\medskip

\heading 5. Conclusions and Remarks
\endheading
\medskip

We have analyzed the process of building $Os(1)_q$ and $su(1,1)_q$
representations in terms of $k$--products of boson (or $q$--boson) operators
by means of the expansion method. The original representation space $\Cal H$
of $Os(1)$ split into $k$ representation spaces $\Cal H_m$, $m=0\ldots k-1$,
just in the same way as at the classical level is described by Luis and
S\'anchez--Soto \cite{16}. We remark that the procedure allows us to
construct any irreducible representation, but some of them take a particularly
simple expression (that only when $k = 2$ can be defined inside the enveloping
algebra). Things can be arranged so that all the isomorphic splitted spaces
 $\Cal H_m\equiv \tilde \Cal H$ support representations equivalent to
the initial one on $\Cal H$. Thus, $\Cal H$ can be written in the form $\Cal H
=\Cal V \otimes \tilde \Cal H$, where $\Cal V$ is a $k$--dimensional space,
whose vectors can be characterized, for instance, by a representation of the
cyclic group of order $k$, $\Cal C_k$.

\medskip

Our study was intended to clarify the representations that $k$--order operators
can give rise. These considerations must be taken into account when they are
used in coherent states (CS) or squeezed states (SS) constructions that have
become so usual lately. In principle, there are different choices for the
vacuum
(corresponding to the spaces $\Cal H_m$) that can give rise to a rich wealth of
coherent states as has been shown in \cite{14}.   For example, in our language
the framework of \cite{14} can be summarized as follows. A standard $q$--CS
$\psi(z)$ can be decomposed using the proyector $P_m$ on each space $ \Cal H_m$
by $\psi_m(z) = P_m\,\psi(z)$. The point is that if a representation of
$Os(1)_q$ is given in the space of functions on $z$, i.e., $\Cal H$, by $\{a\to
z,\  a^+\to D_q(z),\  N\to z/dz\}$, then $\{a\to D(a)\otimes z,\  a^+\to
D(a^+)\otimes D_q(z),\  N\to D(N)\otimes z/dz\}$ will define a representation
in
$\Cal V\otimes \tilde\Cal H$, where $D:\{a,a^+,N\} \to \Cal C_k$ is a certain

asignation on the cyclic group and $\psi_m(z)\in \Cal H_m\equiv \tilde \Cal H$.
This point of view fits well with the discussion of \cite{16} in  the
$q$--deformation context.

\medskip

Multiboson CS have been sudied in \cite{24} concerning their squeezing
properties. When  $q$ is complex, the squeezing has been examined in \cite{12}
in a way that can be readily extended to the multiboson case. Other aplications
of the multiboson formalism about the $q$--deformed Jaynes--Cumming model
is given in \cite{25}, or in the multimode model of Schumaker--Caves in
quantum optics by \cite{26}.

\bigskip
\medskip

\noindent {\bf Acknowledgements.}
\medskip
This work has been partially supported by a DGICYT project (PB91--0196)
from the Ministerio de Educaci\'on y Ciencia de Espa\~na.
\bigskip
\medskip

\noindent {\bf References.}

\medskip
\eightpoint

\ref
\no[{1}]
\book Lie Groups, Lie Algebras and Some of Their Applications
\by H. Gilmore
\yr 1974
\bookinfo Wiley, New York
\endref

\ref
\no[{2}]
\by V.\, G.\, Drinfeld
\paper Quantum Groups
\jour Proceedings of the International Congress of Mathematics,
MRSI Berkeley
\yr 1986
\pages 798
\endref

\ref
\no[{3}]
\by M. Jimbo
\jour Lett. Math. Phys.
\yr 1985
\vol 10
\pages 63

\moreref
\paper II
\yr 1986
\vol 11
\pages 247
\endref

\ref
\no[{4}]
\by T.L.\, Curtright
\jour in {\sl Physics and Geometry,} L--L.\, Chau and W.\,
Nahm, eds., Plenum, New York, 1990
\endref

\ref
\no[{5}]
\by T.L.\, Curtright and C.K.\, Zachos
\jour Phys. Lett. B
\vol 243
\yr 1990
\pages 237
\endref

\ref
\no[{6}]
\by D.\, B.\, Fairlie
\jour J. Phys. A: Math. Gen.
\vol 23
\yr 1990
\pages L183
\endref

\ref
\no[{7}]
\by T.L.\, Curtright, G.I.\, Ghandour and C.K.\, Zachos
\jour J. Math. Phys.
\vol 32
\yr 1991
\pages 676
\endref

\ref
\no[{8}]
\by C.K.\, Zachos
\jour Contemp. Math.
\vol 134
\yr 1992
\pages 351
\endref

\ref
\no[{9}]
\by A.\, Ballesteros and J.\, Negro
\jour J. Phys. A: Math. Gen.
\vol 25
\yr1992
\pages 5945
\endref

\ref
\no[{10}]
\by A.\, Ballesteros and J.\, Negro
\jour {\sl ``Proceedings of the XIX International
Colloquium on Group Theoretical Methods in Physics"}. Anales de F\1sica,
Monograf\1as, Vol. I. M.O., M.S. y J.M.G. (Eds.) CIEMAT/RSEF, Madrid
\yr 1993
\pages 153
\endref

\ref
\no[{11}]
\by J.\, Katriel, A.I.\, Solomon
\jour J.\, Phys.\, A: Math.\, Gen.
\vol 24
\yr 1991
\pages 2093
\endref

\ref
\no[{12}]
\by E. Celeghini, M. Raseti and G. Vitiello
\jour Phys. Rev. Lett.
\vol 66
\yr 1991
\pages 2056
\endref

\ref
\no[{13}]
\by Zhe Chang
\jour J. Math. Phys.
\vol 33
\yr 1992
\pages 3172
\endref

\ref
\no[{14}]
\by Le--Man Kuang, Fa--Bo Wang and Gao--Jian Zeng
\jour Phys. Lett.
\vol A176
\yr 1993
\pages 1
\endref

\ref
\no[{15}]
\by R.A.\, Brandt and O.W.\, Greenberg
\jour J. Math. Phys.
\vol 10
\yr 1969
\pages 1168
\endref

\ref
\no[{16}]
\by A.\, Luis and L.L.\, S\'anchez--Soto
\jour J. Phys. A: Math. Gen.
\vol 24
\yr1991
\pages 2083
\endref

\ref
\no[{17}]
\by P.P.\, Kulish and E.V.\, Damashinski
\jour J.\, Phys.\, A: Math.\, Gen.
\vol 23
\yr 1990
\pages L415
\endref

\ref
\no[{18}]
\by J.\, Katriel, A.I.\, Solomon
\jour J.\, Phys.\, A: Math.\, Gen.
\vol 23
\yr 1991
\pages L1209
\endref

\ref
\no[{19}]
\by L.C.\, Biedenharn
\jour J.\, Phys.\, A: Math.\, Gen.
\vol 22
\yr 1989
\pages L873
\endref

\ref
\no[{20}]
\by A.J.\, Macfarlane
\jour J.\, Phys.\, A: Math. Gen.
\vol 22
\yr 1989
\pages 4581
\endref

\ref
\no[{21}]
\by G. Rideau
\jour Lett.\,Math.\, Phys.\,
\vol 24
\yr 1992
\pages 147
\endref

\ref
\no[{22}]
\by S. Chaturvedi and V. Srinivasan
\jour Phys.\, Rev.\, A
\vol 44
\yr 1991
\pages 8020
\endref

\ref
\no[{23}]
\by T.\, Holstein and H.\, Primakoff
\jour Phys.\, Rev.\,
\vol 58
\yr 1940
\pages 1048
\endref

\ref
\no[{24}]
\by J. Katriel, A.I. Solomon, G. D'Ariano and M. Rasetti
\jour Phys.\, Rev.\, D
\vol 34
\yr 1986
\pages 2332
\endref

\ref
\no[{25}]
\by M. Chaichian, D. Ellinas and P. Kulish
\jour Phys.\, Rev.\, Lett.
\vol 65
\yr 1990
\pages 980
\endref

\ref
\no[{26}]
\by J. Katriel and A.I. Solomon
\jour Quantum Opt.
\vol 1
\yr 1989
\pages 85
\endref

\vfill
\end

\ref
\no[{}]
\by Le--Man Kuang and Fa--Bo Wang
\jour Phys. Lett.
\vol A173
\yr 1993
\pages 221
\endref

\vfill

\pagebreak

\end